# Model of destruction of montmorillonite crystal structure in a microwave field

Makarov V.N.[†,1], Kanygina O.N.[1]

[1]Orenburg State University, Victory Ave. 13, 460018, Orenburg, Russia

[†]e-mail: makarsvet13@gmail.com

We address amorphization of montmorillonite crystal structure. Powdered montmorillonite samples with an effective particle diameter D ≤ 630 μm were treated by a microwave field with frequency 2.45 GHz and power 750 W for ten minutes in different environments. The first sample was treated in an air environment, the second one in a humid environment. The third sample was a ceramic mass with 10% concentration of mixing water (with respect to the total mass). The model describing amorphization is based on results of X-ray analysis. Activation energies were evaluated for covalent and ionic bounds of oxygen ions, hydroxyl groups, and silicon (aluminum) ions. It is shown that the amorphization is carried out in four stages. In fractions of the energy consumption of the last stage the first one consumes 7÷9 %, the second one takes 13÷15%, and the third one takes 49÷59%.

## 1. Introduction

High frequency electromagnetic field treatment is a new approach to modifications of structure of non-organic materials [1-3]. Structure changes of nature-occurring alumosilicates are of a special interest as alumosilicates are widely considered to be a perspective raw material for production of new functional materials via microwave field treatment [4].

Recently montmorillonite-containing and montmorillonite clays were used as raw products for new fictional materials. For dioctahedral aluminuos smectites the corresponding chemical formula reads:

$$(Al_{2-y}Mg_y)(Si_{4-x}Al_x)O_{10}(OH)_2 E_{x+y} \cdot nH_2O, \text{ if } y>x$$

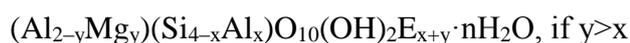

then the smectite is called montmorillonite. Its particles are nature-occurring two-dimensional nanoparticles sensitive to structure changes which makes them suitable for various realizations of structure evolution. Montmorillonite unite cell with an iron ion occupying one of oxygen octahedra (which is a common feature for Orenburg region clays) is presented on Fig. 1 [5]:

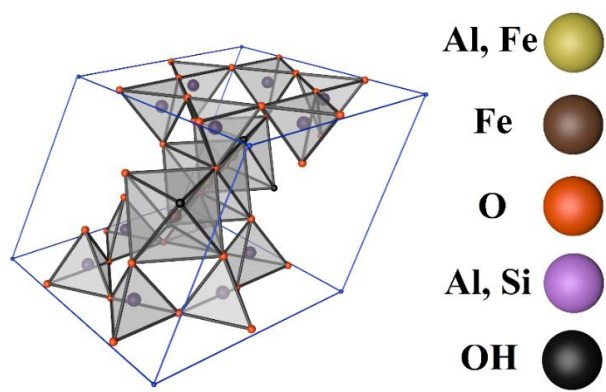

Fig. 1. The unit cell of montmorillonite.

Aluminum and iron ions are located in centers of oxygen octahedrons; aluminum and silicon ions are located in centers of oxygen tetrahedrons. At vertices of oxygen octahedra, a hydroxyl group may be located.

Adsorption, interplanar, and lattice (the hydroxyl group in the lattice) waters are of a special interest for study of montmorillonite structure evolution. Results of the previous thermodynamic studies point to sophisticated interactions between montmorillonite particles and water which are indicated by three endothermic effects associated with the extraction of water of various types. Adsorption water molecules are placed at the boundary of crystal particles and the mechanism responsible for their extraction due to heating, also known as the first endothermic effect, is well-studied both by theoretically and empirically [6]. Interplanar water molecules are placed between neutral layers of the lattice. The correspondent activation mechanism, also known as the second endothermic effect, is also well-studied. For instance, in paper [7] a model was studied which is capable to connect the dehydration temperature, positions of $Al^{3+}$ ions, and their transitions to different parts of oxygen octahedra. The model sucessfuly describes the second endothermic effect.

The lattice water which is a hydroxyl group located within the montmorillonite unit cell is of a special interest. Extraction of hydroxyl groups from montmorillonite unite cells is called the third endothermic effect. In paper [8] an influence of the 10-minute-long microwave field treatment on morphology and phase structure of montmorillonite clays was studied. The particle dispersion effect was found which enhances system adsorption features while keeping the phase structure almost unaffected. The description of enhanced adsorption properties of layered alumosilicates (montmorillonite and kaolin) was given in papers [9,10]. Further study of the third endothermic effect induced by the microwave treatment is relevant, as it allows one to predict montmorillonite clays adsorption features and to design functional materials with enhanced adsorption properties.

In paper [11] it was proven empirically that the degree of montmorillonite amortization is defined by conditions of the high-frequency microwave field treatment. Namely, the microwave treatment of powdered samples resulted in a few effects. Firstly, the intensity of diffraction peaks is decreased. Secondly, interplanar distances of montmorillonite lattice are also decreased due to extraction of water molecules located in between three-layered packages of crystal plane. The present paper presents an attempt to evaluate and to visualize amorphization (destruction) of the montmorillonite crystal structure (the unit cell) after the third endothermic effect.

## 2. Materials and experiments

We use empirical data about destruction of montmorillonite lattice induced by a microwave field treatment for the sake of illustration. A more detailed discussion of these results can be found in [11]. The object of the empirical study was a nature-occurring montmorillonite clay, its chemical composition is discussed in [12]. Powdered samples (P0) were subjected to 10-minute-long microwave treatment (frequency is 2.45 GHz, power is 750 W) in various conditions. In the first case (P1) a 10mm thick powdered sample was placed in a vessel and treated in the air environment. In the second case a sample (P2) was treated in a humid environment (the partial pressure of the water vapor was increased by factor 2). The third sample (P3) was a uniform soft mud consisting of the same clay minerals and 10% (in mass) of mixing water. The diffractogram peaks were measured with DROM-3M apparatus in the refraction regime (Bragg-Brentano geometry) via CuKα radiation with the angular velocity 1 angular degree/min. The particles effective diameter was D ≤ 630 μm. As a result is was established that the montmorillonite unit cell is destroyed and the intensity of the process is defined by the conditions of the experiment. Values of interplanar distances d and the correspondent diffraction peak intensities I are given in Table 1. In the first column the diffractogram line numbers are given (Figure 2).

Table 1. Changes in the crystal structure of montmorillonite induced by the radiation treatment [11]

| No. | hkl | P0 | | P1 | | P2 | | P3 | |
|---|---|---|---|---|---|---|---|---|---|
| | | d, nm | I, % | d, nm | I, % | d, nm | I, % | d, nm | I, % |
| 1 | 001 | 1.22 | 100 | 1.08 | 80 | 0.95 | 70 | 1.41 | 70 |
| 2 | 100 | 0.51 | 20 | 0.51 | 10 | - | 0 | - | 0 |
| 3 | 110, 020 | 0.44 | 80 | 0.44 | 30 | 0.44 | 30 | 0.44 | 30 |
| 4 | 002 | 0.32 | 60 | 0.32 | 40 | 0.31 | 20 | 0.32 | 30 |
| 5 | 200, 130 | 0.26 | 80 | 0.26 | 50 | 0.25 | 30 | 0.25 | 10 |
| 6 | 210 | 0.25 | 40 | - | 0 | - | 0 | - | 0 |
| 7 | 003 | 0.21 | 30 | 0.21 | 10 | - | 0 | 0.21 | 20 |
| 8 | 300, 222 | 0.17 | 70 | 0.17 | 30 | 0.17 | 20 | 0.17 | 30 |
| 9 | 310, 142 | 0.17 | 70 | 0.17 | 50 | - | 0 | 0.17 | 20 |
| 10 | 060, 213 | 0.15 | 80 | 0.15 | 50 | 0.15 | 10 | 0.15 | 20 |

Lined diffraction patterns for all samples are given on Fig. 2. On Figure 2 we present 9 out of 10 diffractogram lines for sample P0. They are required for a study of montmorillonite lattice destruction. For sample P1 8 out 10 lines are given (plane (210) vanishes) that can be observe in the discussed range. For sample P2 5 lines are presented (planes (100), (210), (003), and (310,142) vanish). For sample P3 7 out 10 lines are presented (planes (100) and (210) vanish).

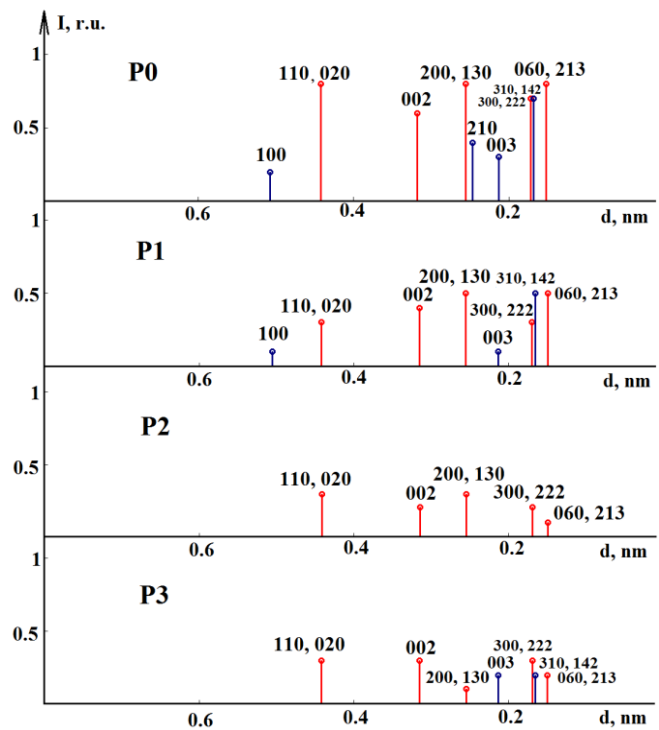

Fig. 2. Diffraction patterns for P0, P1, P2, and P3. After the microwave field treatment the following occurred. For sample P1 the intensity of diffraction maxima decreased and the plane

(210) disappeared; for sample P2 the intensity of diffraction maxima decreased significantly and planes with indices (100), (210), (003), and (310,142) disappeared; for P3 the intensity of diffraction maxima decreased (and become comparable with sample P1) and planes (100) and (210) disappeared.

## 3 Discussion of Results

In order to describe amorphization (destruction) of the montmorillonite unit cell it is required to evaluate the ions bound energy with respect to various crystallographic directions. It is equal to the activation energy required by ions to vacate the unit cell. Calculations were made for a unit cell in an equilibrium state. The evaluation method is based on a calculation of the activation energy for ions located within the crystal plane destroyed after the microwave treatment (see Figure 2). These energies were compared with the empirical data [11]. It should be noted, that we have found a rough estimation of the activation energy. This is due of the fact that the activation energy is a difference between the peak of an activation barrier and the energy of an initial state. The difference between the obtained estimation and the true activation energy can be up to a few orders of magnitude.

Preliminary results were presented in [13]. The activation energy required by a particle to leave the unit cell can be calculated with the linear muffin-tin orbitals in the strong coupling regime [14]. In the present paper an evaluation of the extraction activation energy for hydroxyl groups and other ions of the unit cell is given based on the aforementioned empirical data.

### 3.1. Extraction activation energy for a hydroxyl group and an oxygen atom

Evidence of amorphization can be found in samples P2 and P3 when the crystallographic plane (100) vanishes (Table 1, line №2). At this stage the hydroxyl group and the oxygen cation are removed from oxygen octahedra. Visualization of this amorphization stage presented on Fig. 3. Atoms that cross the plain and leave the crystal due to the high frequency electromagnetic field treatment are marked green.

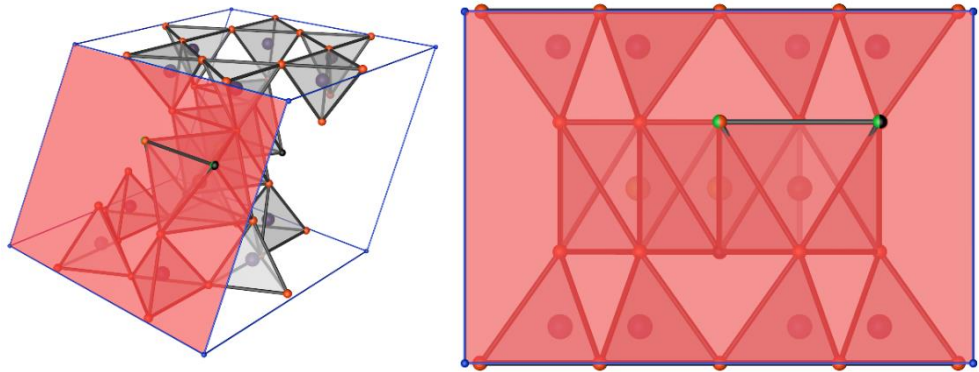

Fig. 3. The crystallographic plane (100) of the unit cell of montmorillonite (two projections)

Extraction activation energies for a hydroxyl group and an oxygen atom defined by an ion placed in the center of the oxygen octahedra:

$$\sum U_{Fe}(100) = \sum U_{Fe-O}(100) + \sum U_{Fe-OH}(100) \qquad (1)$$

or $\sum U_{Al}(100) = \sum U_{Al-O}(100) + \sum U_{Al-OH}(100)$, (2)

where $\sum U_{Fe}(100)$ is the extraction activation energy for the hydroxyl group and the oxygen ion when an *iron* ion is placed in the center of the oxygen octahedron;

$\sum U_{Al}(100)$ is the extraction activation energy for the hydroxyl group and the oxygen ion when an *aluminum* ion occupies the center of the oxygen octahedron.

Extraction activation energy of an oxygen when an *iron* ion occupies the center of the oxygen octahedron is given by the following expression:

$$\sum U_{Fe-O}(100) = \frac{U_{Fe-O(O)} + 2 \cdot E_{O-O(O)} + 2 \cdot E_{OH-O}}{5}, \quad (3)$$

where $E_{O-O(O)}$ is the bound energy between oxygen cations inside the oxygen octahedron;

$E_{OH-O}$ is the bound energy between cations and the hydroxyl group;

$U_{Fe-O(O)}$ is the bound energy between an oxygen cation and the iron ion in an oxygen octahedron.

The bound energy between oxygen cations in an oxygen octahedron can be evaluated with approximate values defined in [15, 16 p. 221], so it reads $E_{O-O(O)} = 234,256$ kJ/mol. The bound energy between an oxygen ion and the hydroxyl group is obtained with approximate values given in [17], and it reads $E_{OH-O} = 20,287$ kJ/mol. The bound energy between an oxygen and the iron ions can be evaluated based on electronegativity [18]:

$$U_{Fe-O(O)} = \frac{1}{2} Z_{(O)} \cdot e \cdot N_A \cdot (\chi_O - \chi_{Fe})^2 = 740,662 \text{ kJ/mol}, \quad (4)$$

here $N_a$ is the Avogadro constant,

e is the electron charge,

$Z_{(O)}$ is the number of nearest neighbors,

$\chi_O = 3,335$ is the oxygen electronegativity,

$\chi_{Fe} = 1,735$ is the iron electronegativity.

The data was used to evaluate the iron activation energy via (3). The energy is $\sum U_{Fe-O}(100) = 249,950$ kJ/mol.

If an aluminum atom occupies the center of an oxygen octahedra, then the extraction activation energy for an oxygen ion is given by the following expression:

$$\sum U_{Al-O}(100) = \frac{U_{Al-O(O)} + 2 \cdot E_{O-O(O)} + 2 \cdot E_{OH-O}}{5}, \quad (5)$$

here $U_{Al-O(O)}$ is the bound energy between oxygen and aluminum ions in an oxygen octahedron.

The bound energy between an oxygen cation and the aluminum ion in the oxygen octahedron can be evaluated via electronegativity[18]:

$$U_{Al-O(O)} = \frac{1}{2} Z_{(O)} \cdot e \cdot N_A \cdot (\chi_O - \chi_{Al})^2 = 831,227 \text{ kJ/mol}, \quad (6)$$

her $\chi_{Al} = 1,640$ is the iron electronegativity. Thus,

$\sum U_{Al-O}(100) = 268,063$ kJ/mol.

If a iron ion is located in the center of an oxygen octahedron, the extraction activation energy for a hydroxyl group reads:

$$\sum U_{Fe-OH}(100) = \frac{E_{Fe-OH} + 4 \cdot E_{OH-O}}{5} \quad (7)$$

here $E_{Fe-OH}$ is the bound energy between the hydroxyl group and the iron ion. The bound energy between the hydroxyl group and the iron ion can be obtained with the approximation data

[15, 19-21], so the energy reads $E_{Fe-OH} = 186,942$ kJ/mol. This allows one to obtain the energy given by (7) which reads $\sum U_{Fe-OH}(100) = 53,618$ kJ/mol.

If an aluminum ion is located in the center of an oxygen octahedron, the extraction activation energy for a thehydroxyl group is given by the following expression:

$$\sum U_{Al-OH}(100) = \frac{E_{Al-OH} + 4 \cdot E_{OH-O}}{5} \quad (8)$$

here $E_{Al-OH}$ is the bound energy between the hydroxyl group and the aluminum ion. The bound energy between the hydroxyl group and the aluminum ion can be found via approximate data [15, 19, 20], thus it reads $E_{Al-OH} = 209,800$ kJ/mol. With these values the energy (8) can ber evaluated and reads $\sum U_{Al-OH}(100) = 58,190$ kJ/mol. Finally, extraction activation energies for a hydroxyl group and an oxygen cation in case when an iron ion is located the center of an oxygen octahedron is given by (1) and reads $\sum U_{Fe}(100) = 303,568$ kJ/mol; in case when an aluminum ion occupies the place the energy is given by (2) and reads $\sum U_{Al}(100) = 326,253$ kJ/mol.

In a similar way a semi-quantitative calculations can be performed for bounds then are destroyed within plains (100), (210), (003). When the activation energy for a hydroxyl group placed within (003) plain, it was assumed that $E_{OH-OH} \approx E_{OH-O}$ as the hydrogen bound can be neglected within the studied setup.

### 3.2. Intensity dynamics for lines №9 (310, 142) in P2

During the discussed stage of the experiment oxygen cations are extracted from a tetrahedral ring within plain (142). During this stage plains (310) lose oxygen cations and cations of tetrahedral rings (aluminum and silicon) which results in strong crystal amorphization. Visualization of this stage presented on Fig. 4. These results are in agreement with the theoretical results, but they do not fit the extraction activation condition for ions in octahedral silicon packages [22].

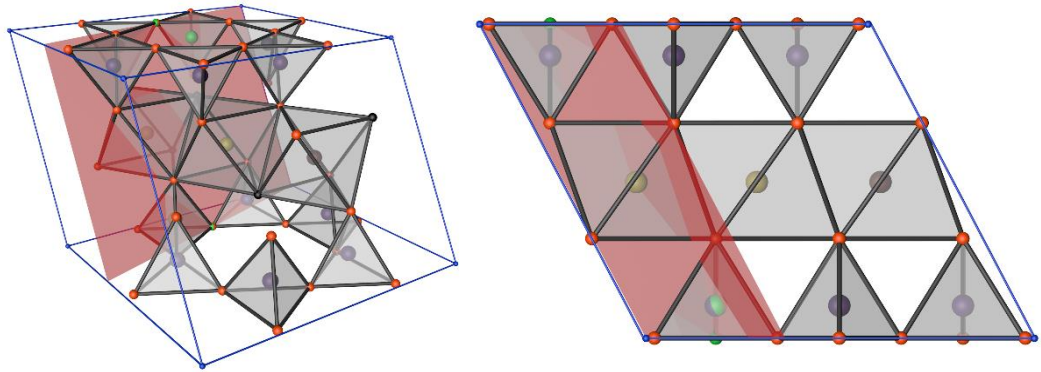

Fig. 4. The crystallographic plane (310) for the unit cell of montmorillonite (two projections)

Let us study the plain (310).
Within an oxygen tetrahedron the extraction activation energy for oxygen cations and an ion reads:

$$\sum U^{Si}(310) = \sum U_{Si-O}(310) + \sum U_{Si}(310), \quad (9)$$

or $$\sum U^{Al}(310) = \sum U_{Al-O}(310) + \sum U_{Al}(310), \quad (10)$$

or $$\sum U^{Si,Al}(310) = \sum U_{Si-O,Al-O}(310) + \sum U_{Si,Al}(310). \quad (11)$$

Formulae (9) and (10) describe maximal and minimal values of the energy correspondingly.

When a *silicon* ion is located in the center of an oxygen tetrahedron the extraction activation energy for oxygen cations reads:

$$\sum U_{Si-O}(310) = 2 \cdot \frac{2 \cdot U_{Si-O(T)} + 6 \cdot E_{O-O(T)}}{8} = 482,235 \text{ kJ/mol.} \quad (12)$$

When an *aluminum* ion is located in the center of an oxygen tetrahedron the extraction activation energy for oxygen cations is:

$$\sum U_{Al-O}(310) = 2 \cdot \frac{2 \cdot U_{Al-O(T)} + 6 \cdot E_{O-O(T)}}{8} = 609,826 \text{ kJ/mol.} \quad (13)$$

Finally, when an *aluminum* ion occupies the center of one tetrahedron and a *silicon* ion occupies another, then the extraction activation energy of oxygen cations reads:

$$\sum U_{Si-O,Al-O}(310) = 2 \cdot \frac{U_{Si-O(T)} + U_{Al-O(T)} + 6 \cdot E_{O-O(T)}}{8} = 546,030 \text{ kJ/mol.} \quad (14)$$

Calculations of the extraction activation energy should account not only for ion bounds, but also for covalent bounds. Ratios [23] between metallic, ion, and covalent bounds that holds between the oxygen and the other elements were used. Thus, the covalent bound reads:

$$U^C_{Si(T)} = \frac{100 \cdot U_{Si-O(T)}}{22,93} = 1303,838 \text{ kJ/mol,} \quad (15)$$

$$U^C_{Al(T)} = \frac{100 \cdot U_{Al-O(T)}}{30,19} = 1835,546 \text{ kJ/mol,} \quad (16)$$

here $U^C_{Si(T)}$ is the covalent bound energy for the silicon ion inside an oxygen tetrahedron.

$U^C_{Al(T)}$ is the covalent bound energy for the aluminum ion in an oxygen tetrahedron.

The extraction activation energies for the (silicon or aluminum) ion inside an oxygen tetrahedron are given by the following formulae:

$$\sum U_{Si}(310) = 2 \cdot U^C_{Si(T)} = 2607,676 \text{ kJ/mol;} \quad (17)$$

$$\sum U_{Al}(310) = 2 \cdot U^C_{Al(T)} = 3671,092 \text{ kJ/mol;} \quad (18)$$

$$\sum U_{Si,Al}(310) = U^C_{Si(T)} + U^C_{Al(T)} = 3139,384 \text{ kJ/mol.} \quad (19)$$

Extraction activation energies for an oxygen cation and the (silicon or aluminum) ion inside an oxygen tetrahedron for formulae (9), (10), (11) read:

$$\sum U^{Si}(310) = 3089,911 \text{ kJ/mol,}$$

or $\sum U^{Al}(310) = 4280,918$ kJ/mol,

or $\sum U^{Si,Al}(310) = 3685,414$ kJ/mol.

Extraction activation energies for the oxygen cation and the (silicon or aluminum) ion inside an oxygen tetrahedron for formulae (9), (10), and (11) read:

$\sum U^{Si}(310,142) = 3331,029$ kJ/mol is the ion extraction activation energy if a silicon ion is placed in an oxygen tetrahedron.

$\sum U^{Al}(310,142) = 4585,831$ kJ/mol is the ion extraction activation energy if an aluminum ion is placed in an oxygen tetrahedron.

Maximal and minimal values of the ion extraction activation energy for plains (310, 142) defined by positions of ions within the montmorillonite lattice.

## 3.3 Maximal and minimal values of the ion extraction activation energy for plains (310, 142) depend on positions of ions within the montmorillonite lattice

Values of ion extraction activation energies for the studied plains presented in Table 2.

Table 2. Activation energies of ion extraction from crystallographic plane

| Indices of the plane, (hkl) | Ions leaving the unit cell | The activation energy of ions, kJ/mol |
|---|---|---|
| 100 | O, OH | 304 ÷ 326 |
| 210 | O | 482 ÷ 610 |
| 003 | O, OH | 1948 ÷ 2260 |
| 310, 142 | O, (Si, Al) | 3331 ÷ 4586 |

Results show that ion extraction activation energies for plains (100) and (210) are very close to each other. Desctruction of plains (003) and (310,142) requires an increase of the energy by an order of magnitude.

According to the X-ray diffraction data and to the numerical calculations the amorphization takes four stages. Visualization of these stages is given on Fig. 5.

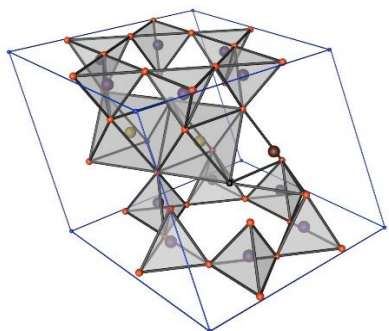

Fig. 5.a. The first stage of amorphization of the montmorillonite cell; ion activation at (100)

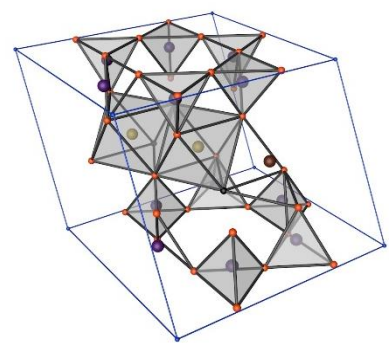

Fig. 5.b. The second stage of the amorphization of the montmorillonite cell; ion activation at (210)

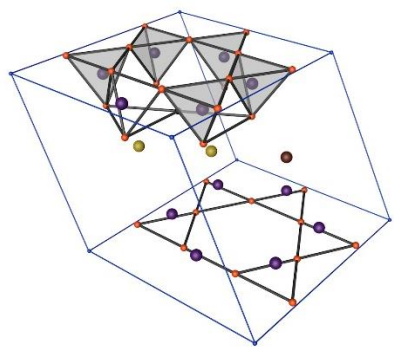

Fig. 5.c. The third stage of the amorphization of the montmorillonite cell; ion activation at (003)

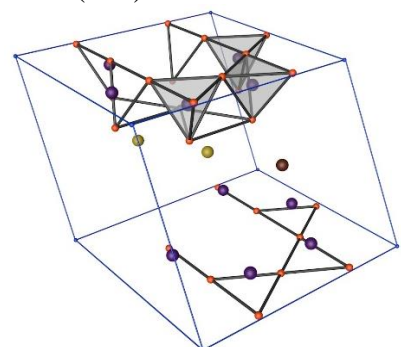

Fig. 5.d. The fourth stage of the amorphization of the montmorillonite cell; ion activation at (310, 142)

## 4. Conclusions

Montmorillonite unit cell amorphization induced by high-frequency electromagnetic field treatment can be split in four stages.

During the first stage the oxygen octahedron in the plain (100) is destroyed due to extraction of oxygen ions and hydroxyl group ions (Fig. 5.a).

During the second stage oxygen tetrahedra are destroyed in plains (210) due to extraction of oxygen ions (Fig. 5.b). Because ion extraction activation energies for ions place in plains (100) and (210) are similar, first and second stages of the amorphization are equally probable. For sample P1 the ion extraction occured in the plain (210) while the diffraction maximum intensity (100) decreased down to 10%. For sample P3 ions were extracted from plains (100) and (210) which proves that ion extraction activation energies have comparable values for these plains.

During the third stage an intensive lattice destruction takes place along plains (003) due to extraction of oxygen ions and hydroxyl groups (Fig. 5.c). This results in a destruction of oxygen tetrahedra and octahedra.

During the fourth stage the lattice is destroyed along plains (310,142) because of the extraction of oxygen and aluminum or silicon ions (Fig.5.d). The fourth stage in the most energy intensive as it requires destruction of covalent bound.